\documentclass[conference]{IEEEtran}
\usepackage{amsmath,amsfonts}
\usepackage{algorithmic}
\usepackage{algorithm}
\usepackage[caption=false,font=scriptsize]{subfig}
\usepackage{graphicx}
\usepackage{cite}
\usepackage{flushend}
\usepackage{color}
\IEEEoverridecommandlockouts

\begin{document}

\title{Low-Complexity Detection of Multiple Preambles in the Presence of Mobility and Delay Spread\vspace{-3mm}}

\author{Sandesh Rao Mattu, Beyza Dabak, Venkatesh Khammammetti, and Robert Calderbank,~\IEEEmembership{Fellow,~IEEE}\\
Department of Electrical and Computer Engineering, Duke University, USA \vspace{-15mm}
\thanks{The partnership between the Duke, IIT Delhi, and IISC Bengaluru is supported by the US National Science Foundation under grant 2342690 and by the India Department of Science and Technology. The Duke team is also supported in part by the National Science Foundation under grant 2148212 and is supported in part by funds from federal agency and industry partners as specified in the Resilient \& Intelligent NextG Systems (RINGS) program. The Duke team is supported in part by the Air Force Office of Scientific Research under grants FA 8750-20-2-0504 and FA 9550-23-1-0249.

This work may be submitted to the IEEE for possible publication. Copyright
may be transferred without notice, after which this version may no longer be
accessible.}
}



\maketitle

\begin{abstract}
Current wireless infrastructure is optimized to support applications such as music/video streaming and internet browsing, where information flows from the base station to the user. This paper anticipates the emergence of applications such as distributed machine learning and automated driving, resulting in a shift of engineering focus from downlink to uplink as users become significant sources of data. The current paradigm of scheduling users on reserved uplink resources (grants) is not able to deal efficiently with unpredictable traffic patterns. As a result, Release 15, 3GPP introduced the 2-step RACH as a mechanism to enable grant-free (random) initial access. The first of the two steps is preamble detection in a RACH slot, and in this paper we describe a very low-complexity algorithm for simultaneous detection of multiple preambles in the presence of mobility and delay spread. We provide a pathway to standards adoption by choosing Zadoff-Chu (ZC) sequences as preambles, taking advantage of the fact that ZC sequences already appear in 5G standards. We construct preambles by using the discrete Zak transform to pass from a ZC sequence of length $MN$ in the time domain (TD) to a quasi-periodic $M\times N$ array in the delay-Doppler (DD) domain. There are $MN$ quasi-periodic Dirac pulses, each corresponding to a Zak-OTFS carrier waveform, and the ZC preamble is simply the corresponding sum of Zak-OTFS carrier waveforms. We detect multiple preambles in the presence of mobility and delay spread by sampling the received signal on the $M\times N$ period grid in the DD domain. We approach detection as a compressed sensing problem. We represent a preamble as a column of length $M N$ in the DD domain and apply discrete shifts in delay and Doppler to produce a block with $\mathcal{O}(MN)$ columns in the compressed sensing matrix. The superposition of multiple preambles determines a block sparse sum of columns in the sensing matrix. The correlation properties of ZC sequences result in a highly structured compressed sensing matrix, making it possible to identify constituent preambles using One-Step Thresholding (OST), which has complexity $\mathcal{O}(M^3N^3)$. In this paper, we describe an algorithm with complexity that is $\mathcal{O}(M^2N^2)$ in the size of an individual column ($MN$). 
\end{abstract}

\vspace{-5mm}
\section{Introduction}
Current wireless infrastructure is optimized to support applications such as music/video streaming and internet browsing. The engineering objective is a high-throughput downlink that maximizes information flow from the base station to the user \cite{liu2001opportunistic}, \cite{qin2006distributed}, \cite{eryilmaz2007fair} \cite{choi2007opportunistic}, \cite{hou2009theory}, \cite{gopalan2012wireless}. As applications such as distributed machine learning and automated driving emerge, we anticipate that users will become significant sources of data and that engineering focus will shift from the downlink to the uplink. This paper addresses the challenge of providing low-latency and ubiquitous connectivity for the combination of short packets, unpredictable traffic patterns, and high Doppler spreads which is characteristic of Mobile Massive Machine-Type Communications (3MTC) in 6G propagation environments.

The current paradigm of scheduling users on reserved uplink resources (grants) is not able to deal with unpredictable traffic patterns efficiently, and random access at the medium access control (MAC) layer provides a promising alternative. This was recognized by Release 15, 3GPP which introduced the 2-step RACH as a mechanism to enable grant-free (random) initial access. We note that there has been growing interest is supporting massive multiple access using the 2-step RACH procedure \cite{2-step-workitem}, \cite{peralta2021two}, \cite{agostini2024evolution}. The first of the two steps is preamble detection in a RACH slot, and in this paper, we describe a low-complexity algorithm for simultaneous detection of multiple preambles in the presence of mobility and delay spread. We provide a pathway to standards adoption by using Zadoff-Chu (ZC) sequences as preambles, taking advantage of the fact that they are already part of 5G standards. In this paper, we do not consider the second step in the 2-step RACH, which is is uplink data transmission in a PSCH slot identified by a detected preamble. 

Preamble detection is critical as a first step, because if a preamble is not detected, then the pointer to the radio resources used for uplink data transmission is lost. This results in lost data, and it also increases interference with other users. In Section III we describe the detection of multiple preambles in the RACH slot in the presence of mobility and delay spread as a compressed sensing problem. The columns of the compressed sensing matrix are the different preambles shifted in delay and in Doppler. We consider $M$ discrete delay shifts and $N$ discrete Doppler shifts so that the compressed sensing matrix is finite.

A Zak-OTFS carrier waveform is a pulse in the delay-Doppler (DD) domain, that is a quasi-periodic localized function defined by a delay period $\tau_p$ and a Doppler period $\nu_p = 1/ \tau_p$. When viewed in the time-domain (TD) this function is realized as a pulse train modulated by a tone, hence the name pulsone. The time duration ($T$) and bandwidth ($B$) of a pulsone are inversely proportional to the characteristic width of the DD domain pulse along the Doppler axis and the delay axis respectively. The number of non-overlapping DD domain pulses, each spread over an area $1/BT$, is equal to the time-bandwidth product $BT$. The corresponding pulsones are orthogonal to one another, rendering Zak-OTFS an orthogonal modulation that achieves the Nyquist rate. Throughout this paper, we set $M=\tau_p/B$ and $N=\nu_p/T$, so that there are $MN$ quasi-periodic DD domain pulses, each corresponding to a Zak-OTFS carrier waveform (see \cite{bits1}, \cite{bits2} for more details). We choose the delay period to be greater than the delay spread $\tau_{\max}$ and the Doppler period to be greater than the Doppler spread $2\nu_{\max}$, a condition we call the crystallization condition that guarantees the Zak-OTFS modulation is predictable and non-fading.

Section II describes how the discrete Zak transform represents sequences of composite length $MN$ in the time domain (TD) as $M \times N$ arrays in the delay-Doppler (DD) domain. The discrete Zak transform is a unitary transform between periodic sequences with period $MN$ and $M \times N$ quasi-periodic arrays. Inner products in the time domain are preserved after transformation to the DD domain. The ZC sequence maps to a quasi-periodic $M \times N$ array and the ZC preamble is simply the corresponding sum of Zak-OTFS carrier waveforms. We detect multiple preambles after sampling on the $M \times N$ period grid. Each preamble determines a column of length $MN$ in a compressed sensing matrix $\mathbf{A}$. We consider $x$ discrete delay shifts (where $x(\tau_p/M)=\tau_{\max}$) and $2y$ discrete Doppler shifts (where $y(\nu_p/N)= \nu_{\max}$) for a total of $xy = MN (\tau_{\max} \nu_{\max})$ columns, which determine a block with $\mathcal{O}(MN)$ columns in the compressed sensing matrix $\mathbf{A}$ that is indexed by the preamble. In Section III we describe how correlation properties of ZC sequences result in highly structured compressed sensing matrices. Note that the correspondence described in Section II is not specific to ZC sequences, that there are other families of spreading sequences with good correlation properties in the time domain, and that each family determines a highly structured compressed sensing matrix in the same way.


When we transmit multiple preambles in a RACH slot, then sample on the $M\times N$ grid, we produce a signal $\mathbf{y}$ that is a block sparse sum of columns of a highly structured compressed sensing matrix $\mathbf{A}$. The number of blocks involved in the sum is simply the number of active preambles. The coefficients in the sum specify the Input / Output (I/O) relation for a sampled Zak-OTFS system. We are detecting multiple preambles from the sampled signal, and we emphasize that the receiver has no knowledge of the physical channel other than the worst case delay and Doppler spreads. In fact, the simulations presented in Section IV are performed for the most challenging situation, which is the combination of unresolvable paths and high channel spreads, and the fractional delay and Doppler is significant. In previous work \cite{ost_zc} we used knowledge of the I/O relation to support uplink data transmission using Zak-OTFS, but here we only focus on preamble detection. We demonstrate that it is possible to detect multiple preambles in the presence of mobility and delay spread. By contrast, the performance of the TD matched filter detector degrades severely in the presence of Doppler \cite{sat_rand_acc}.

The classical theory of compressed sensing emphasizes Gaussian sensing matrices and reconstruction via convex minimization \cite{donoho2006}, \cite{candes2006}. Highly structured compressed sensing matrices, such as those defined by preambles, open the door to reconstruction algorithms such as One-Step Thresholding (OST) that have much lower complexity \cite{OST}. In previous work, we used OST \cite{ost_zc} to detect multiple preambles in the presence of mobility and delay spread. However, the compressed sensing matrix has $M N$ rows and $\mathcal{O}((M N )^2)$ columns and the complexity of OST is cubic in the size of an individual column.

The structure of our compressed sensing matrix is similar to that of a second compressed sensing matrix that appears in Gaussian random access \cite{chirrup}. The columns of this matrix are 2nd-order Reed Muller codewords, and the compressed sensing problem is to recover the constituents of a real-valued linear combination of codewords. The entries of each column are indexed by binary vectors $\mathbf{v}$ of length $m$, the columns are indexed by quadratic forms $Q$, and the $v$th entry of column Q is simply $Q(v)$. Let $N = 2^m$. CHIRRUP is a non-linear algorithm with complexity $KN\log_2(N)$ that recovers a superposition of $K$ codewords of length $N$. Note that complexity depends on the length of a column rather than the number of columns.

ZC sequences of length $MN$ in the TD are indexed by roots $u$ that are coprime to $MN$. The exponent of the $n$th entry is $- un(n+1)/2$ which is quadratic in $n$. This quadratic structure is present in the DD domain after we apply the discrete Zak transform to obtain the ZC preamble, and it remains present when the preamble is shifted in delay and Doppler. The property that quadratic structure is preserved by delay and Doppler shifts is particular to Zak-OTFS modulation, and it is this property that enables low-complexity detection of multiple preambles in the presence of mobility and delay spread. Section III describes the chirp detection algorithm and Section IV presents simulation results that are representative of 6G propagation environments.

\section{The Geometry of the Zak Transform}
We refer the interested reader to \cite{ost_zc} for a complete description of the Zak-OTFS system model used in this paper. In this section, we will show that the inner product in the delay-Doppler (DD) domain is equal to the inner product in the time-domain. Consider two time-domain sequences $\mathbf{x}[n]$ and $\mathbf{y}[n]$ with $n=0, 1, \cdots, MN-1$. Let the corresponding DD domain representations be denoted by $\mathbf{X}^{(\mathrm{dd})}[k, l]$ and $\mathbf{Y}^{(\mathrm{dd})}[k, l]$, respectively, where $k=0, 1, \cdots, M-1$ and $l = 0, 1, \cdots, N-1$. The inner product in the DD domain is given by
\begin{align}
    Z^{(\mathrm{dd})} = \sum_{k=0}^{M-1} \sum_{l=0}^{N-1} \mathbf{X}^{(\mathrm{dd})}[k, l] \bar{\mathbf{Y}}^{(\mathrm{dd})}[k, l],
    \label{inner_prod}
\end{align}
where $\bar{\mathbf{Y}}^{(\mathrm{dd})}[k, l]$ is the complex conjugate of $\mathbf{Y}^{(\mathrm{dd})}[k, l]$. The discrete Zak transform (DZT) \cite{dzt} expresses the DD domain representation as
\begin{align}
    \mathbf{X}^{(\mathrm{dd})}[k, l] = \frac{1}{\sqrt{N}} \sum_{n=0}^{N-1} \mathbf{x}[k+nM]\xi_N^{-ln},
    \label{dzt}
\end{align}
where $\xi_N$ is a primitive $N$th root of unity. Substituting \eqref{dzt} in \eqref{inner_prod},
\begin{align}
    Z^{(\mathrm{dd})} &= \sum_{k=0}^{M-1} \sum_{l=0}^{N-1}\frac{1}{\sqrt{N}}\sum_{n=0}^{N-1}\mathbf{x}[k+nM]\xi^{-ln}_N \times \nonumber \\
    &\hspace{10mm}\frac{1}{\sqrt{N}}\sum_{m=0}^{N-1}\bar{\mathbf{y}}[k+mM]\xi_N^{lm} \nonumber \\
    &= \frac{1}{N}\sum_{k=0}^{M-1}\sum_{l=0}^{N-1}\sum_{n=0}^{N-1}\sum_{m=0}^{N-1}\mathbf{x}[k+nM] \times \nonumber \\
    &\hspace{10mm}\bar{\mathbf{y}}[k+mM]\xi_N^{l(m-n)} \nonumber \\
    &= \frac{1}{N}\sum_{k=0}^{M-1}\sum_{n=0}^{N-1}\sum_{m=0}^{N-1}\mathbf{x}[k+nM] \times \nonumber \\
    &\hspace{10mm}\bar{\mathbf{y}}[k+mM]\sum_{l=0}^{N-1}\xi_N^{l(m-n)}.
    \label{derivation_1}
\end{align}
The sum
\begin{align}
    \sum_{l=0}^{N-1}\xi_N^{l(m-n)} = \begin{cases}
        N, \quad \mathrm{if }\ m = n \mod N \\
        0, \quad \hspace{1mm}\mathrm{otherwise}
    \end{cases}.
\end{align}
Since $0 \leq m, n \leq N-1$, the sum is non-zero only if $m=n$. Substituting $m=n$ in \eqref{derivation_1}, we have
\begin{align}
    Z^{(\mathrm{dd})} &= \sum_{k=0}^{M-1}\sum_{n=0}^{N-1} \mathbf{x}[k+nM] \bar{\mathbf{y}}[k+nM] \nonumber \\
    &= \sum_{j=0}^{MN-1} \mathbf{x}[j] \bar{\mathbf{y}}[j] = Z^{(\mathrm{t})}.
    \label{derivation_2}
\end{align}
The right-hand side of the expression in \eqref{derivation_2} is the inner product of the two sequences $\mathbf{x}[n]$ and $\mathbf{y}[n]$ in the time-domain. The discrete Zak transform preserves inner products.

\section{Chirp Detection}

A Zadoff-Chu (ZC) sequence $\mathbf{x}_{u}[n]$ with root $u$ and composite length $MN$ is given by
\begin{equation}
    \mathbf{x}_{u}[n]=\xi_{M N}^{-u n(n+1)/2} \quad \text { where } n=0,1, \ldots, M N-1,
\end{equation}
where $\xi_{MN}$ is a primitive $MN$th root of unity.
Given $0<a<MN$, with $a$ coprime to $M$ and $N$, we shift $\mathbf{x}_u[n]$ in time to obtain  $\mathbf{x}_u[n+a]$ where the indices $n+a$ are read modulo $MN$.
Then
\begin{align}
    \mathbf{x}_{u}[n+a]=\xi_{M N}^{-u (n+a)(n+a+1)/2}. 
\end{align}
Pointwise multiplication yields
\begin{align}
    \mathbf{z}_{u}[n] = \mathbf{x}_{u}[n]\bar{\mathbf{x}}_{u}[n+a]&=\xi_{M N}^{-u n(n+1)/2}\xi_{M N}^{u (n+a)(n+a+1)/2} \nonumber \\
    &=\xi_{MN}^{ua(a+1)/2} \xi_{MN}^{uan}.
    \label{td_multiply}
\end{align}
Taking the $MN$-point Fast Fourier transform (FFT) of \eqref{td_multiply} gives
\begin{align}
    \mathcal{F}_{MN}\{\mathbf{z}_{u}[n]\} = \xi_{MN}^{ua(a+1)/2} \delta(f - ua),
\end{align}
where $\delta(\cdot)$ is the Kronecker delta function. The peak of the frequency response is at $ua$ and $u$ can be determined since $a$ is known. A simple frequency counter therefore serves to detect $u$. 
This is the idea behind chirp detection \cite{chirp_reconstruction}.

\subsection{Chirp Detection for Doubly Selective Channels}
Chirp detection is able to identify the root $u$ on a Gaussian channel, and we now describe how to extend this method to doubly selective channels.

The first step is to convert \eqref{td_multiply} to the DD domain using the DZT\footnote{
DZT maps sequences of length $MN$ in the TD to $M \times N$ arrays in the DD domain. The DZT of a sequence $\mathbf{x}[n], n=0, 1, \cdots, MN-1,$ is a matrix $\mathbf{X}^{(\mathrm{dd})}[k, l], k=0, 1, \cdots, M-1, l = 0, 1, \cdots, N-1$ where $\mathbf{X}^{(\mathrm{dd})}[k, l] = \frac{1}{\sqrt{N}}\sum_{n=0}^{N-1} \mathbf{x}[k+nM]\xi_N^{-ln}$.}
which we denote by $\mathcal{Z}$. In the DD domain,
\begin{align}
    \mathbf{Z}_{u}^{(\mathrm{dd})}[k, l] &= \mathcal{Z}\{\mathbf{z}_u[n]\} \nonumber \\
    &= \frac{1}{\sqrt{N}} \sum_{n=0}^{N-1} \mathbf{z}_u[k + nM] \xi_{N}^{-ln} \nonumber \\
    &= \frac{1}{\sqrt{N}} \sum_{n=0}^{N-1} \xi_{MN}^{ua(a+1)/2} \xi_{MN}^{ua(k+nM)} \xi_{N}^{-ln} \nonumber \\
    &= \frac{1}{\sqrt{N}} \xi_{MN}^{ua(a+1)/2} \xi_{MN}^{uak} \sum_{n=0}^{N-1} \xi_{N}^{(ua-l)n}.
    \label{dd_proc}
\end{align}
The sum
\begin{align}
    \sum_{n=0}^{N-1} \xi_{N}^{(ua-l)n} = \begin{cases}
        N \quad \text{ if } l = ua \mod N \\
        0 \quad \text{ otherwise}
    \end{cases}.
    \label{sum_n}
\end{align}
Substituting \eqref{sum_n} in \eqref{dd_proc}, we have
\begin{align}
    \mathbf{Z}_{u}^{(\mathrm{dd})}[k, l] = \begin{cases}
        \sqrt{N} \xi_{MN}^{ua(a+1)/2} \xi_{MN}^{uak} \quad \text{ if } l = ua \mod N \\
        0 \quad\quad\quad\quad\quad\quad\quad\quad\ \  \text{ otherwise }
    \end{cases},
\end{align}
which implies that $\mathbf{Z}_{u}^{(\mathrm{dd})}[k, l]$ is supported on the line $l = ua \mod N$. 
Since $a$ is coprime to $N$, there is a unique inverse $a^{-1}$ and
\begin{align}
    u = la^{-1} \mod N
    \label{l_line}
\end{align}
represents the set of possible $u$'s with $0 < u < MN$.

The second step is to convert \eqref{td_multiply} to the time-frequency (TF) domain by first converting the vector $\mathbf{z}_u[n]$ into a time-delay matrix $\mathbf{Z}_u \in \mathbb{C}^{M \times N}$ and taking an $M$-point FFT along the delay axis, i.e.,
\begin{align}
    \mathbf{Z}_u^{(\mathrm{tf})}[k, l] &= \mathcal{F}_{M}\{\mathbf{Z}_u\} \nonumber \\
    &= \frac{1}{\sqrt{M}} \sum_{n=0}^{M-1} \mathbf{z}_u[l+nN]\xi_{M}^{-kn} \nonumber \\
    &= \frac{1}{\sqrt{M}} \sum_{n=0}^{M-1} \xi_{MN}^{ua(a+1)/2} \xi_{MN}^{ua(l+nN)} \xi_{M}^{-kn} \nonumber \\
    &= \frac{1}{\sqrt{M}} \xi_{MN}^{ua(a+1)/2} \xi_{MN}^{ual} \sum_{n=0}^{M-1} \xi_{M}^{(ua-k)n}.
    \label{tf_proc}
\end{align}
The sum
\begin{align}
    \sum_{n=0}^{M-1} \xi_{M}^{(ua-k)n} = \begin{cases}
        M \quad \text{ if } k = ua \mod M \\
        0 \quad \text{ otherwise}
    \end{cases}.
    \label{sum_m}
\end{align}
Substituting \eqref{sum_m} in \eqref{dd_proc}, we have
\begin{align}
    \mathbf{Z}_{u}^{(\mathrm{tf})}[k, l] = \begin{cases}
        \sqrt{M} \xi_{MN}^{ua(a+1)/2} \xi_{MN}^{ual} \quad \text{ if } k = ua \mod M \\
        0 \quad\quad\quad\quad\quad\quad\quad\quad\ \  \text{ otherwise }
    \end{cases},
\end{align}
which implies that $\mathbf{Z}_{u}^{(\mathrm{tf})}[k, l]$ is also supported on the line given by $k = ua \mod M$. 
Since $a$ is coprime to $M$, there is a unique inverse $a^{-1}$ and
\begin{align}
    u = ka^{-1} \mod M
    \label{k_line}
\end{align}
represents another set of possible $u$'s with $0 < u < MN$.

Equations \eqref{l_line} and \eqref{k_line} represent two lines with slopes $N$ and $M$, respectively. 
These lines meet at a unique point, and the point of intersection identifies the root $u$ (see Fig. \ref{fig:alg_pictorial}).

\subsection{One User, Multiple Paths}
\begin{figure*}
    \subfloat[{ DD domain}]{\includegraphics[width=0.32\linewidth]{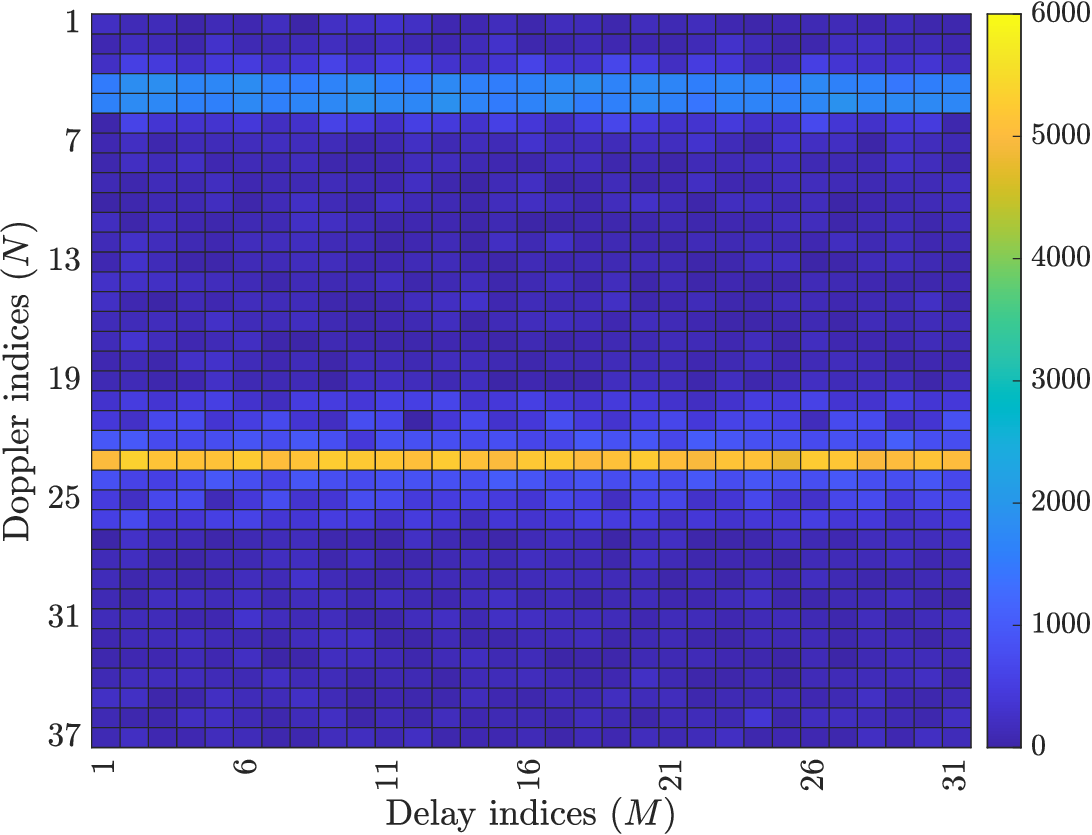}\label{fig:l_line}}
    \hfill
    \subfloat[{ TF domain}]{\includegraphics[width=0.32\linewidth]{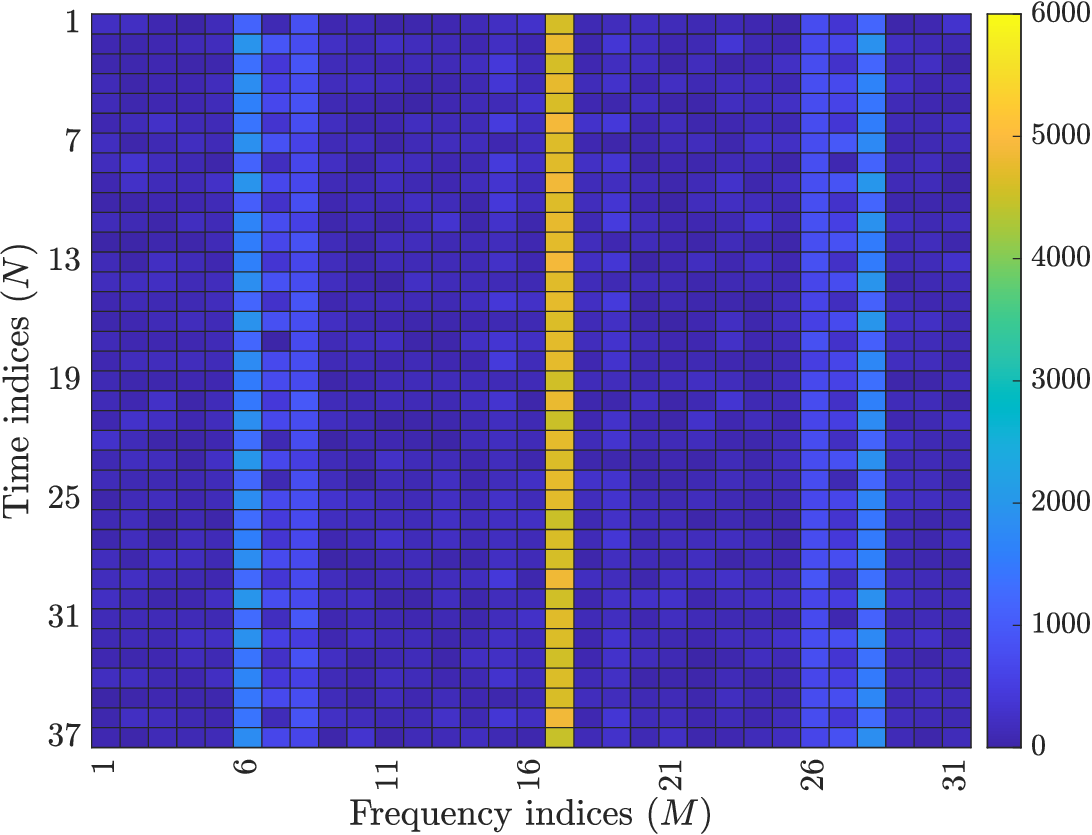}\label{fig:k_line}}
    \hfill
    \subfloat[{ Intersection of two lines}]{\includegraphics[width=0.345\linewidth]{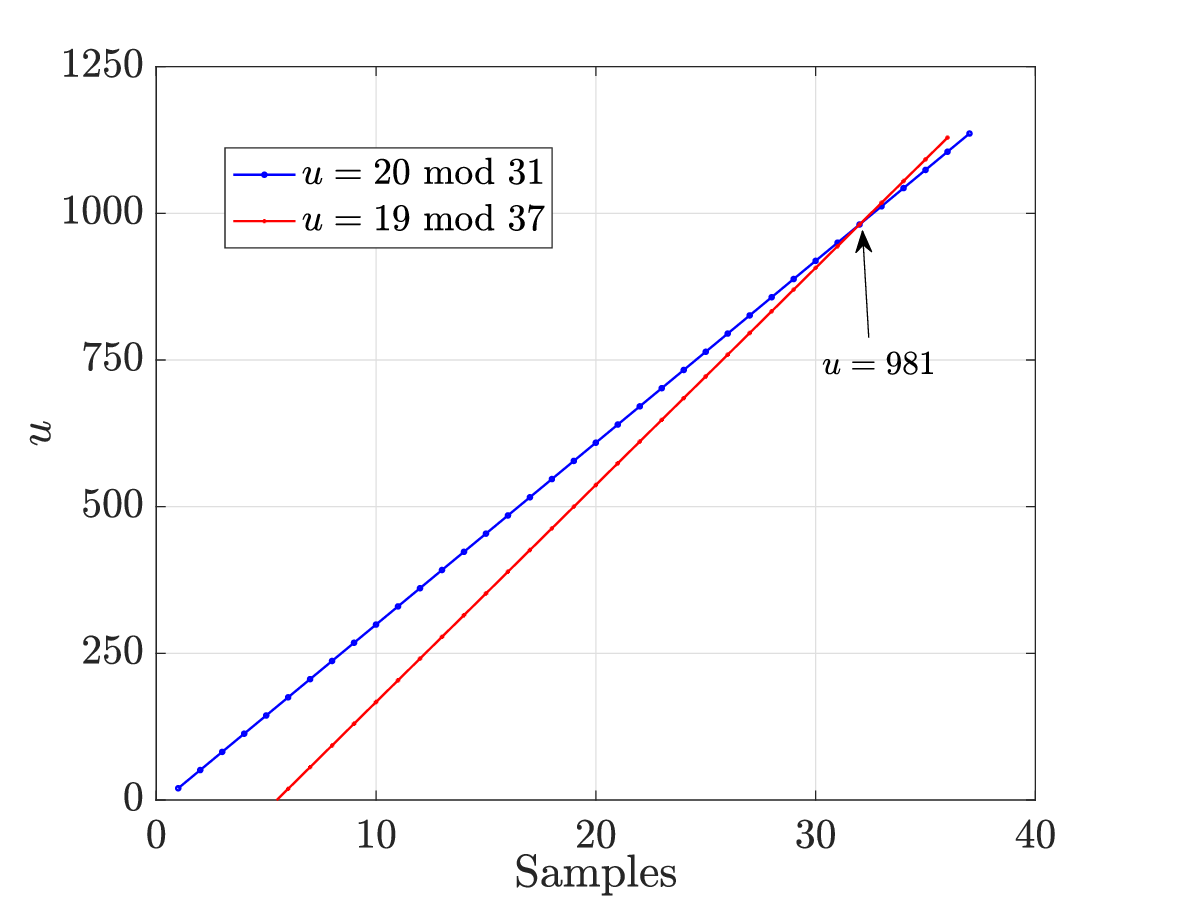}\label{fig:intersect_k_l}}
    \caption{The figures show how the proposed Algorithm \ref{alg:one_user} works for one user. $M=31, N=37$ for Zak-OTFS grid with Doppler period $\nu_p = 30 $ kHz. Veh-A channel, RRC pulse shaping filter ($\beta_\tau = \beta_\nu = 0.6$), $\nu_{\max} = 815$ Hz. User SNR is taken to be $20$ dB. The root chosen by the user is $u=981$ and shift is $a=7$. From (a), $l=22 (= 981 \times 7 \mod 37)$ (in zero indexing) corresponds to the maximum amplitude and from (b), $k=16 (= 981 \times 7 \mod 31)$. Inverse of $7$ under $N=37$ is $16$ and under $M=31$ is $9$. Therefore $u = 22 \times 16 \mod 37 = 19 \mod 37$ (from (a)) and $u = 16 \times 9 \mod 31 = 20 \mod 31$ (from (b)). These lines are plotted in (c) and their intersection is at $u=981$.}
    \label{fig:alg_pictorial}
\end{figure*}

The idea behind the algorithm is the following. We start with one preamble transmitted by the user and receive a sum of multiple shifts of the preamble due to the channel. 
We shift and conjugate this sum, then perform pointwise multiplication to obtain a sum of diagonal (self) terms from individual paths and cross terms from pairs of paths. The correlation properties of ZC sequences and their shifts imply that the cross terms are small, and Algorithm \ref{alg:one_user} treats these cross terms as noise\footnote{Although the derivation points to the line $l = ua \mod N$ (or $k = ua \mod M$), due to the effect of noise and channel, it is difficult to pick the correct column (or row) index. Therefore, we sum the absolute values for each column (or row) and pick the index corresponding to the largest sum among the summed values (see Fig. \ref{fig:alg_pictorial}).}.

\begin{algorithm}
    \caption{Detection of one user}
    \label{alg:one_user}
    \begin{algorithmic}[1]
        \STATE  \textbf{Input:} A received time domain signal $\mathbf{y}[n]$, shift $a$
        \STATE \textbf{Output:} Detected root $u$
        \STATE Delay the received time domain signal by $a$ to obtain $\mathbf{y}[n+a]$
        \STATE Compute $\mathbf{z}_a[n] = \mathbf{y}[n]\bar{\mathbf{y}}[n+a]$
        \STATE Obtain $\mathbf{Z}_a^{(\mathrm{dd})}[k, l]$ from $\mathbf{z}_a[n]$ using DZT
        \STATE Compute absolute sum along the rows in $\mathbf{Z}_a^{(\mathrm{dd})}[k, l]$, i.e., $\mathbf{s}_a^{(\mathrm{dd})}[l] = \sum_{k = 0}^{M-1} \vert \mathbf{Z}_a^{(\mathrm{dd})}[k, l] \vert$
        \STATE Obtain the index with the highest absolute sum, i.e., \\$l' = \underset{l}{\text{ arg max }} \mathbf{s}_a^{(\mathrm{dd})}[l]$
        \STATE Obtain the first set $\mathcal{U}_1$ of $u$'s, using \eqref{l_line}
        \STATE Obtain $\mathbf{Z}_a^{(\mathrm{tf})}[k, l]$ from $\mathbf{z}_a[n]$ using $M$-point FFT
        \STATE Compute absolute sum along the columns in $\mathbf{Z}_a^{(\mathrm{tf})}[k, l]$, i.e., $\mathbf{s}_a^{(\mathrm{tf})}[k] = \sum_{l = 0}^{N-1} \vert \mathbf{Z}_a^{(\mathrm{tf})}[k, l] \vert$
        \STATE Obtain the index with the highest absolute sum, i.e., \\$k' = \underset{k}{\text{ arg max }} \mathbf{s}_a^{(\mathrm{tf})}[k]$
        \STATE Obtain the second set $\mathcal{U}_2$ of $u$'s, using \eqref{k_line}
        \STATE Compute $\mathcal{U}_1 \cap \mathcal{U}_2$ to obtain $u$
        \RETURN $u$
    \end{algorithmic}
\end{algorithm}

\subsection{$K$ Users, Multiple Paths}
We now describe how to determine the vector $u = [u_1 \ u_2 \ \cdots \ u_{K-1}]$ of roots when we receive $K$ preambles.

The received sequence (without the effects of noise and channel) is expressed as
\begin{align}
    \mathbf{x}_{\mathbf{u}}[n] = \sum_{i=0}^{K-1} \xi_{MN}^{-u_in(n+1)/2}.
\end{align}
Delaying $\mathbf{x}_{\mathbf{u}}[n]$ by $a$,
\begin{align}
    \mathbf{x}_{\mathbf{u}}[n + a] = \sum_{i=0}^{K-1} \xi_{MN}^{-u_i(n+a)(n+a+1)/2},
\end{align}
and multiplying its conjugate with $\mathbf{x}_{\mathbf{u}}[n]$ results in
\begin{align}
    \mathbf{z}_{\mathbf{u}}[n] &= \mathbf{x}_{\mathbf{u}}[n]\bar{\mathbf{x}}_{\mathbf{u}}[n+a] \nonumber \\
    &= \sum_{i=0}^{K-1}\xi_{M N}^{-u_i n(n+1)/2} \sum_{j=0}^{K-1} \xi_{M N}^{u_j (n+a)(n+a+1)/2} \nonumber \\
    &= \sum_{i=0}^{K-1}\xi_{M N}^{-u_i n(n+1)/2} \Bigg(\xi_{M N}^{u_i (n+a)(n+a+1)/2} + \nonumber \\
    &\hspace{20mm}\sum_{j=0, j\neq i}^{K-1} \xi_{M N}^{u_j (n+a)(n+a+1)/2}\Bigg) \nonumber \\
    &= \sum_{i=0}^{K-1}\xi_{MN}^{u_ia(a+1)/2} \xi_{MN}^{u_ian} + \nonumber \\
    & \hspace{5mm}\sum_{i=0}^{K-1}\sum_{j=0, j\neq i}^{K-1}\xi_{M N}^{-u_i n(n+1)/2} \xi_{M N}^{u_j (n+a)(n+a+1)/2}.
    \label{many_users_td}
\end{align}
The right hand side of \eqref{many_users_td} has two sums, namely $S_1 = \sum_{i=0}^{K-1}\xi_{MN}^{u_ia(a+1)/2} \xi_{MN}^{u_ian}$ and $S_2 = \sum_{i=0}^{K-1}\sum_{j=0, j\neq i}^{K-1}\xi_{M N}^{-u_i n(n+1)/2} \xi_{M N}^{u_j (n+a)(n+a+1)/2}$. For $S_1$ a similar approach as described in Algorithm $\ref{alg:one_user}$ can be used for obtaining $u_i$'s with the change that instead of picking one maximum index (Steps 7 and 11 in Algorithm \ref{alg:one_user}) $K$ maximum indices need to be picked. However, $S_2$ introduces significant cross term ($i \neq j$) interference and the above-proposed approach no longer works. Therefore we modify Algorithm \ref{alg:one_user} to better suit this setting.

\textit{Remark on OST:} One step thresholding (OST) \cite{OST} is a compressed sensing algorithm used to solve a sparse recovery problem. The idea here is to construct an observations matrix $\mathbf{A}$ with columns constructed from ZC sequences with certain delay and Doppler shifts. At the receiver $\mathbf{A}^{+}$ (hermitian of $\mathbf{A}$) is multiplied with the received vector $\mathbf{y}$ and a maximum energy detection is performed for determining the transmitted ZC sequence(s). The reader is referred to \cite[Sec. VI]{ost_zc} for more details.

\begin{algorithm}
    \caption{Detection of $K$ users}
    \label{alg:k_users}
    \begin{algorithmic}[1]
        \STATE  \textbf{Input:} A received time domain signal $\mathbf{y}[n]$, corresponding DD domain vector $\mathbf{y}^{(\mathrm{dd})}[n]$, shifts $\mathbf{a} = [a_0 \ a_1 \ \cdots \ a_{I-1}]$, observations matrix $\mathbf{A}$
        \STATE \textbf{Output:} Detected vector $\hat{\mathbf{u}} = [\hat{u}_0 \ \hat{u}_1 \ \cdots \ \hat{u}_{K-1}]$
        \FOR{$a_i \in \mathbf{a}$}
            \STATE Delay the received time domain signal by $a_i$ to obtain $\mathbf{y}[n+a_i]$
            \STATE Compute $\mathbf{z}_{a_i}[n] = \mathbf{y}[n]\bar{\mathbf{y}}[n+a_i]$
            \STATE Obtain $\mathbf{Z}_{a_i}^{(\mathrm{dd})}[k, l]$ from $\mathbf{z}_{a_i}[n]$ using DZT
            \STATE Compute absolute sum along the rows in $\mathbf{Z}_{a_i}^{(\mathrm{dd})}[k, l]$, i.e., $\mathbf{s}_{a_i}^{(\mathrm{dd})}[l] = \sum_{k = 0}^{M-1} \vert \mathbf{Z}_{a_i}^{(\mathrm{dd})}[k, l] \vert$
            \STATE Obtain $K$ indices with highest absolute sums, i.e., \\$\mathcal{L} = \{l \subset \{0, 1, \cdots, N-1\} \mid \mathbf{s}_{a_i}^{(\mathrm{dd})}[l] > \mathbf{s}_{a_i}^{(\mathrm{dd})}[l'] \ \forall \ l' \not\in \mathcal{L} \text{ and } \vert \mathcal{L} \vert = K\}$ 
            \FOR{$l \in \mathcal{L}$}
                \STATE Compute $u$'s using \eqref{l_line} and add them to the set $\mathcal{U}_{a_i}$
            \ENDFOR
            \STATE Obtain $\mathbf{Z}_{a_i}^{(\mathrm{tf})}[k, l]$ from $\mathbf{z}_{a_i}[n]$ using $M$-point FFT
            \STATE Compute absolute sum along the columns in $\mathbf{Z}_{a_i}^{(\mathrm{tf})}[k, l]$, i.e., $\mathbf{s}_{a_i}^{(\mathrm{tf})}[k] = \sum_{l = 0}^{N-1} \vert \mathbf{Z}_{a_i}^{(\mathrm{tf})}[k, l] \vert$
            \STATE Obtain $K$ indices with highest absolute sums, i.e., \\$\mathcal{K} = \{k \subset \{0, 1, \cdots, M-1\} \mid \mathbf{s}_{a_i}^{(\mathrm{tf})}[k] > \mathbf{s}_{a_i}^{(\mathrm{tf})}[k'] \ \forall \ k' \not\in \mathcal{K} \text{ and } \vert \mathcal{K} \vert = K\}$ 
            \FOR{$k \in \mathcal{K}$}
                \STATE Compute $u$'s using \eqref{k_line} and add them to the set $\mathcal{V}_{a_i}$
            \ENDFOR
        \ENDFOR
        \STATE Compute $\mathcal{U} = \{\mathcal{U}_{a_i} \cap \mathcal{V}_{a_j}, i, j = 0, 1, \cdots, I-1\}$
        \STATE For each $u \in \mathcal{U}$, pick corresponding columns from $\mathbf{A}$ (see \cite[Eq. (64)]{ost_zc}) to get the matrix $\mathbf{A}'$
        \STATE Compute $\mathbf{f} = \mathbf{A'}^{+}\mathbf{y}^{(\mathrm{dd})}$ (OST algorithm)
        \STATE Sum energies of entries in $\mathbf{f}$ corresponding to each root $u \in \mathcal{U}$
        \STATE Pick $\hat{\mathbf{u}} = [\hat{u}_0 \ \hat{u}_1 \ \cdots \ \hat{u}_{K-1}]$ corresponding to maximum sum energies
        \RETURN Return $\hat{\mathbf{u}}$
    \end{algorithmic}
\end{algorithm}

Algorithm 2 identifies individual roots from a sum of $K$ preambles.
Instead of using a single shift $a$ as previously proposed, we use a vector of $I$ shifts $\mathbf{a}$. For each shift $a_i \in \mathbf{a}$, the delayed conjugated time domain signal is multiplied by the unchanged time domain signal and converted to the DD domain where the absolute row sum is computed (steps 4-7 in Algorithms \ref{alg:one_user} and \ref{alg:k_users}) to obtain $\mathbf{s}_{a_i}^{(\mathrm{dd})}$. Next, from $\mathbf{s}_{a_i}^{(\mathrm{dd})}$, $K$ indices with highest absolute sums are picked to obtain $\mathcal{L}$. For each $l \in \mathcal{L}$, a set of $u$'s is obtained and stored in $\mathcal{U}_{a_i}$. A similar procedure is repeated in the TF domain to obtain another set of $u$'s given by $\mathcal{V}_{a_i}$ (steps 12-17 in Algorithm \ref{alg:k_users}). 
After obtaining $\mathcal{U}_{a_i}$ and $\mathcal{V}_{a_i}$, the intersection between each pair of $\mathcal{U}_{a_i}$ and $\mathcal{V}_{a_j}$ for $i, j = 0, 1, \cdots, I-1$ provides the value of $u$ which is stored in $\mathcal{U}$. From the observations matrix $\mathbf{A}$\footnote{The observations matrix $\mathbf{A}$ is constructed from the ZC sequences in the DD domain. Each column is a ZC sequence translated by delay and Doppler shift and a set of columns corresponds to one ZC sequence (see \cite[Eq. (64)]{ost_zc}). Therefore, the number of columns in the matrix $\mathbf{A}$ is the number of ZC sequences times the number of delay-Doppler translates for each ZC sequence and the number of rows is $MN$.}, columns corresponding to ZC sequence with roots $u \in \mathcal{U}$ are extracted into the matrix $\mathbf{A}'$ (step 21). Next, the OST algorithm is carried out in DD domain using $\mathbf{A}'$ and $\mathbf{y}^{(\mathrm{dd})}$ (which is the DD domain vector corresponding to $\mathbf{y}$) to obtain $\hat{\mathbf{u}} = [\hat{u}_0 \ \hat{u}_1 \ \cdots \ \hat{u}_{K-1}]$, the detected roots (steps 22 and 23).

\begin{table}[t]
    \centering
    \caption{Power-delay profile of Veh-A channel model}
    \begin{tabular}{|c|c|c|c|c|c|c|}
         \hline
         Path number ($i$) & 1 & 2 & 3 & 4 & 5 & 6 \\
         \hline
         $\tau_i (\mu s)$ & 0 & 0.31 & 0.71 & 1.09 & 1.73 & 2.51 \\
         \hline
         Relative power ($p_i$) dB & 0 & -1 & -9 & -10 & -15 & -20 \\
         \hline
    \end{tabular}
    \label{tab:veh_a}
\end{table}

\section{Simulation Results}
In this section, we present the results obtained for single user and $K$ users. For all the simulations, we employ Zak-OTFS with the number of delay bins $M=31$, the number of Doppler bins $N=37$, the Doppler period $\nu_p = 30$ kHz, and delay period $\tau_p = 1/\nu_p = 33.33 \ \mu$s. This corresponds to bandwidth $B = M\nu_p = 930$ kHz and duration $T = N\tau_p = 1.23$ ms. 
We consider root-raised cosine (RRC) shaping filters given by
\begin{align}
    w(\tau, \nu) = \sqrt{BT} \mathrm{rrc}_{\beta_\tau}(B\tau)\mathrm{rrc}_{\beta_\nu}(T\nu),
\end{align}
where $0 \leq \beta_\tau, \beta_\nu \leq 1$ and
\begin{align}
    \mathrm{rrc}_{\beta} = \frac{\sin(\pi x (1-\beta)) + 4\beta x\cos(\pi x(1+\beta))}{\pi x(1-(4\beta x)^2)}
\end{align}
 with parameters $\beta_\tau = \beta_\nu = 0.6$. 
 We consider the six-path Veh-A channel \cite{veh_a} with the power-delay profile listed in Table \ref{tab:veh_a}. The channel path gains ($h_i$) are normalized such that $\sum_{i=0}^{P-1}\vert h_i\vert^2 = 1$, and the channel path Doppler shift for the $i$th path is randomly generated using $\nu_i = \nu_{\max} \cos(\theta_i)$, where $\theta_i \sim \mathrm{U}[-\pi, \pi)$, $\mathrm{U}$ is the uniform distribution and $\nu_{\max} = 815$ Hz. ZC sequences of length $MN$ are used with $G = 1024$ roots (of the $1147$ possible). We present the results first for one user, then for $K$ users. We focus on the probability of misdetection ($1-P_d$), i.e., the probability that each detected root is not present in the set of transmitted roots (we do not map the root back to the user).

\subsection{One User, Multiple Paths}
\begin{figure}
    \centering
    \includegraphics[width=\linewidth]{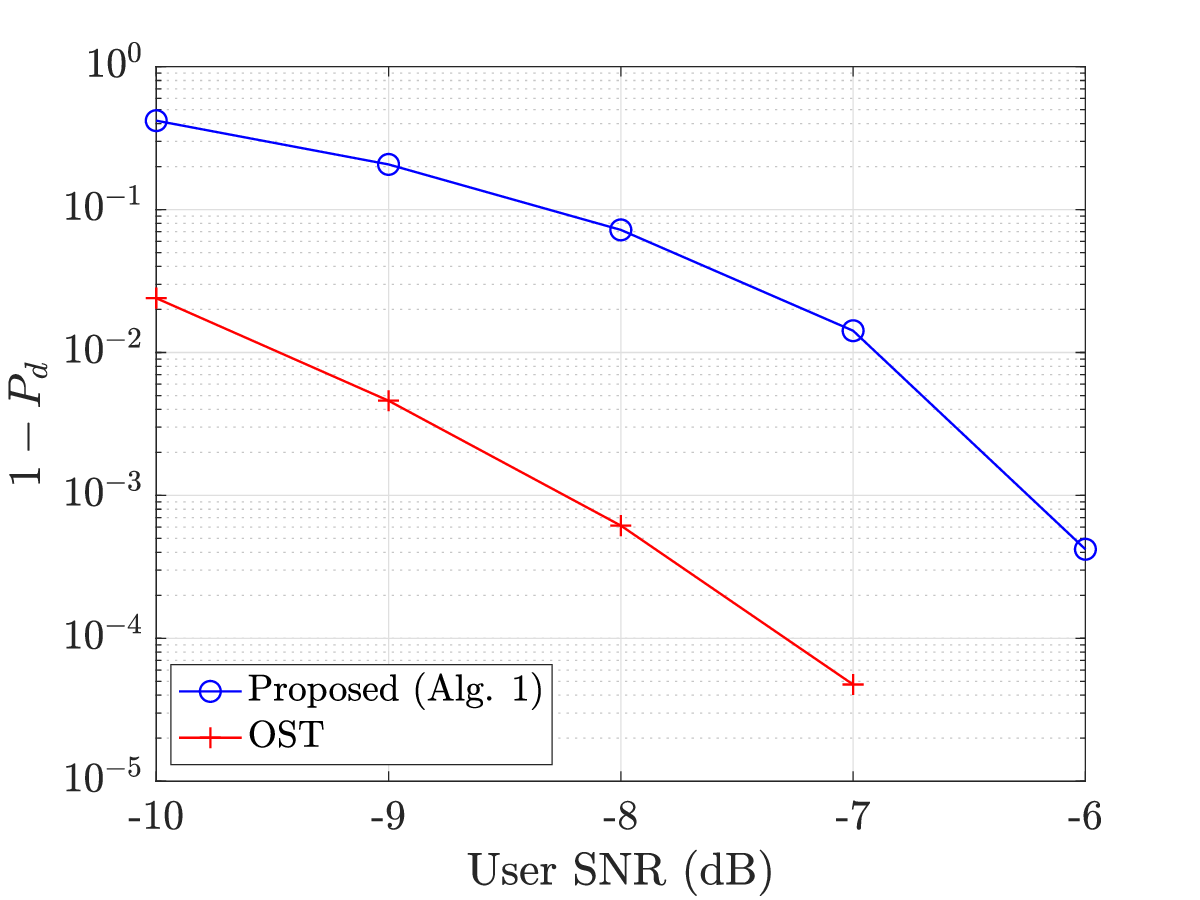}
    \caption{Probability of missed detection ($1-P_d$) as a function of user SNR for one user. $M=31, N=37$ for a Zak-OTFS grid with Doppler period $\nu_p = 30$ kHz. Veh-A channel, RRC pulse shaping filter ($\beta_\tau = \beta_\nu = 0.6$), $\nu_{\max} = 815$ Hz. The performance of the proposed approach (Alg. 1) is close to that of OST differing only by 2 dB.}
    \vspace{-2mm}
    \label{fig:misdet_1_user}
\end{figure}
Figure \ref{fig:misdet_1_user} shows the probability of missed detection as a function of user SNR for one user. The performance of OST receiver \cite{OST} is also added for comparison. It is seen that the performance of the proposed receiver (Algorithm \ref{alg:one_user}) is close to that of OST worse only by 2 dB. However, the performance is achieved at a complexity that is linear in $MN$ compared to cubic in $MN$ of OST.
\begin{table}[t]
    \centering
    \caption{Complexity involved in operations for one user (Algorithm \ref{alg:one_user})}
    \begin{tabular}{c|c}
        \textbf{Operation} & \textbf{Complexity} \\
        \hline
        $z[n] = \mathbf{y}[n]\bar{\mathbf{y}}[n+a]$ & $\mathcal{O}(MN)$ \\ 
        \hline
        DZT ($\mathbf{Z}_a^{(\mathrm{dd})}[k, l]$) & $\mathcal{O}(N \log N)$ \\
        \hline
        Absolute row sum ($\mathbf{s}_a^{(\mathrm{dd})}[l]$) & $\mathcal{O}(M)$ \\
        \hline
        arg max & $\mathcal{O}(N)$ \\
        \hline
        Computing $u$'s in \eqref{l_line} & $\mathcal{O}(M)$  \\
        \hline
        FFT ($\mathbf{Z}_a^{(\mathrm{tf})}[k, l]$) & $\mathcal{O}(M \log M)$ \\
        \hline
        Absolute row sum ($\mathbf{s}_a^{(\mathrm{tf})}[k]$) & $\mathcal{O}(N)$ \\
        \hline
        arg max & $\mathcal{O}(M)$ \\
        \hline
        Computing $u$'s in \eqref{k_line} & $\mathcal{O}(N)$ \\
        \hline
        Finding $u$ & $\mathcal{O}(\max\{M, N\} \log(\max\{M, N\}))$ \\
    \end{tabular}
    \label{tab:one_user}
\end{table}

\begin{table}[t]
    \centering
    \caption{Complexity involved in operations for $K$ users (Algorithm \ref{alg:k_users}) 
    }
    \begin{tabular}{c|c}
        \textbf{Operation} & \textbf{Complexity} \\
        \hline
        $z[n] = \mathbf{y}[n]\bar{\mathbf{y}}[n+a]$ & $\mathcal{O}(IMN)$ \\ 
        \hline
        DZT ($\mathbf{Z}_a^{(\mathrm{dd})}[k, l]$) & $I\mathcal{O}(N \log N)$ \\
        \hline
        Absolute row sum ($\mathbf{s}_a^{(\mathrm{dd})}[l]$) & $\mathcal{O}(IM)$ \\
        \hline
        Top $K$ indices & $I\mathcal{O}(N)$ \\
        \hline
        Computing $u$'s in \eqref{l_line} & $\mathcal{O}(IKM)$ \\
        \hline
        FFT ($\mathbf{Z}_a^{(\mathrm{tf})}[k, l]$) & $I\mathcal{O}(M \log M)$ \\
        \hline
        Absolute row sum ($\mathbf{s}_a^{(\mathrm{tf})}[k]$) & $\mathcal{O}(IN)$ \\
        \hline
        Top $K$ indices & $I\mathcal{O}(M)$ \\
        \hline
        Computing $u$'s in \eqref{k_line} & $\mathcal{O}(IKN)$ \\
        \hline
        Finding the set of $u$'s ($\mathcal{U}$) & $I^2\mathcal{O}(\max\{M, N\} \log (\max\{M, N\}))$ \\
        \hline
        OST & $\mathcal{O}(I^2K^2(MN)^2)$ \\
        \hline
        Sum energies & $\mathcal{O}(I^2K^2MN)$ \\
        \hline
        Top $K$ indices & $\mathcal{O}(I^2K^2)$ \\
    \end{tabular}
    \label{tab:k_users}
\end{table}


\begin{table}[t]
    \centering
    \caption{Complexity involved in OST for $K$ users \cite{OST} 
    }
    \begin{tabular}{c|c}
        \textbf{Operation} & \textbf{Complexity} \\
        \hline
        OST & $\mathcal{O}(GM^2N^2)\approx \mathcal{O}(M^3N^3)$ \\
        \hline
        Sum energies & $\mathcal{O}(GMN)\approx \mathcal{O}(M^2N^2)$ \\
        \hline
        Top $K$ indices & $\mathcal{O}(G) \approx \mathcal{O}(MN)$ \\
    \end{tabular}
    \label{tab:ost}
\end{table}

\subsubsection{Complexity}
\label{sec:comp_1_user}
The complexity of the proposed algorithm (Algorithm \ref{alg:one_user}) is presented in Table \ref{tab:one_user}. 
From the table, the complexity is of the order of $MN$, i.e., $\mathcal{O}(MN)$. In contrast, the complexity of the OST algorithm (presented in Table \ref{tab:ost}) is of the order of $\mathcal{O}(M^3N^3)$ (assuming $G \approx MN$ which is true since $G=1024$ and $MN = 1147$). The complexity of the proposed Algorithm~\ref{alg:one_user} grows linearly in $MN$ while that of OST grows as cube of $MN$.


\subsection{K Users, Multiple Paths}
\begin{figure}
    \centering
    \includegraphics[width=.95\linewidth]{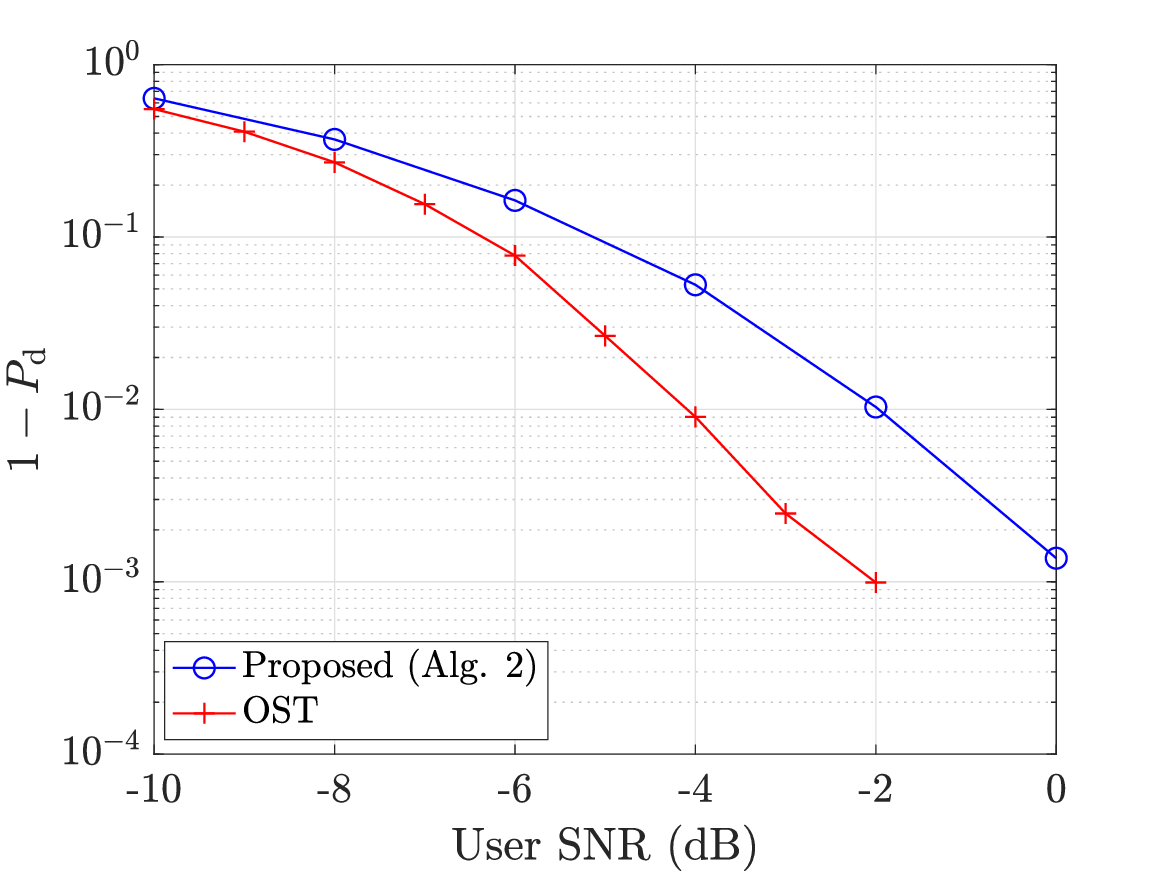}
    \vspace{-3mm}
    \caption{Probability of missed detection ($1-P_d$) as a function of user SNR for $K=5$ active users. $M=31, N=37$ for a Zak-OTFS grid with Doppler period $\nu_p = 30$ KHz. Veh-A channel, RRC pulse shaping filter ($\beta_\tau = \beta_\nu = 0.6$), $\nu_{\max} = 815$ Hz. The performance of the proposed approach (Alg. 2) is close to that of OST differing only by 2 dB.}
    \label{fig:misdet_k_users}
    \vspace{-3mm}
\end{figure}

Figure \ref{fig:misdet_k_users} shows the probability of missed detection as a function of user SNR for $K=5$ simultaneous users. Note that, each user has its own channel matrix and noise realizations and this knowledge is not assumed at the receiver. For construction of the compressed sensing matrix $\mathbf{A}$, only the worst-case delay and Doppler are assumed to be known at the receiver. The performance of OST receiver \cite{OST} is also added for comparison. It is seen that the performance of the proposed receiver (Algorithm \ref{alg:k_users}) is close to that of OST worse only by 2 dB. However, this is obtained at a significantly lower complexity when compared to OST as described below.

\subsubsection{Complexity}
The complexity of the Algorithm \ref{alg:k_users} is presented in Table \ref{tab:k_users}. Since there are $I$ shifts in each step, each complexity computation is multiplied by $I$. 
It is seen that the complexity is quadratic in the number of columns $MN$.\footnote{We note that the complexity is $\mathcal{O}(I^2K^2(MN)^2)$. Since $I$ is chosen during the simulations and $K$ is the number of users, $K$ and $I$ are each $\mathcal{O}(1)$.} As discussed in Section \ref{sec:comp_1_user}, the complexity of OST is cubic in the number of columns implying that the proposed algorithm is computationally efficient by an order of the number of columns.

\section{Conclusion}
Detection of multiple preambles is a key component of the 2-step RACH mechanism introduced in 3GPP, Release 15 to enable grant-free uplink access. We have designed a low-complexity algorithm for the detection of multiple preambles in the presence of mobility and delay spread. Further, we have provided a pathway to standards adoption by choosing Zadoff-Chu (ZC) sequences as preambles, given that ZC sequences already appear in 5G standards. Our algorithm takes advantage of the quadratic structure of ZC sequences. This structure is present when ZC sequences are defined in the time domain, and it remains present in the DD domain after we apply the discrete Zak transform to obtain the ZC preamble. Multiple preamble detection is made possible by the interaction between this quadratic structure and the Zak-OTFS modulation, the fact that the quadratic structure remains present in delay and Doppler shifts of the ZC preamble. 
For the single-user case, the complexity is linear in the time-bandwidth product while for $K$ users it is quadratic in the time-bandwidth product.
\bibliographystyle{IEEEtran}

\bibliography{references}

\end{document}